\documentstyle[prc,aps,epsf,psfig,amssymb]{revtex}

\newcommand \be{\begin{eqnarray}}
\newcommand \ee{\end{eqnarray}}
\begin{document}
%\draft
%\LARGE
%\Large
%\preprint{HEP/123-qed}
\twocolumn[\hsize\textwidth\columnwidth\hsize
          \csname @twocolumnfalse\endcsname
\title{Correlations in Many-Body Systems with Two-time Green's Functions}
\author{H. S. K\"ohler\\}
\address{Physics Department, University of Arizona, Tucson, Arizona
85721}
\author{K. Morawetz}
\address{
Max-Planck-Institute for the Physics of Complex Systems, 
Noethnitzer Str. 38, 01187 Dresden, Germany\\
}
\date{\today}
\maketitle
\begin{abstract}
The Kadanoff-Baym (KB) equations are solved numerically for infinite nuclear
matter.
In particular we calculate correlation energies and correlation times.
Approximating the Green's
functions in the KB collision kernel by the free Green's functions the
Levinson equation is obtained. This approximation 
is valid for weak interactions and/or
low densities. It relates to the
extended quasi-classical approximation for the spectral function. 
%The Levinson correlation
%energy reduces for large times to a second order Born approximation 
%for the energy.  
Comparing the Levinson, Born and KB calculations allows for
an estimate of higher order spectral corrections to the correlations.
A decrease in binding energy is reported due to spectral correlations
and off-shell parts in the reduced density matrix.
\end{abstract}

\pacs{05.20.Dd,05.60.+w,21.65.+f,25.70.-z}

\vskip2pc]

\section{Introduction\protect\\} 

The quantum Kadanoff-Baym equations (KB) \cite{kad62} describe the time-evolution
of the two-time (one-particle) Green's functions $G({\bf p},t,t')$. Imposing
various approximations they have played an
important role in the past developing corrections to the classical
Boltzmann equation such
as memory-effect and damping. With some restrictions it 
is now however feasible to solve
these equations numerically without approximations.

We like to emphasize the following differences between the Boltzmann and
KB-equations.
In the quasi - classical Markovian Boltzmann-equation
correlations are explicitly neglected. The spectral
functions are given by the quasi-particle approximation.
The kinetic energy is conserved in each binary collision, while the
correlation energy is neglected. 
The equilibrium density-distribution (in momentum space)
is a Fermi-distribution. Oppositely, in the KB-equations 
correlations are carried by
the two-time Green's functions. The total energy including the interaction 
potential energy is conserved.  The reduced density distribution is given by a
frequency integral of a spectral-function of non-zero width and a distribution 
function. At equilibrium this distribution-function is again given by a
Fermi-distribution.

The various approximations of the KB-equations  differ essentially by the
reduction schemes of the two - time Green's function to
the reduced density matrix or even to the quasiparticle distribution.
That this reduction is nontrivial can be traced to the fact that
the Fermi and quasi-particle distributions decrease exponentially with
energy while the reduced density matrix possesses power tails. 
In this paper we address some common schemes from the numerical point
of view.

Numerical results of the quantum KB-equations
already have been compared in the past with the classical Markovian
dynamics as well as with other frequently used approximations
\cite{dan84,hsk96}.
The study was made in particular with reference to collisions between heavy
ions usually studied with the Boltzmann-equation or the BUU,VUU etc
versions thereof. The general conclusion of the study was that
quantum-effects reduce relaxation-rates by a factor as small as one
half. This fact can be interpreted as being due to the
pole renormalization or wave
function renormalization \cite{LSM97}.

Since the first numerical applications of the KB-equations 
by Danielewicz \cite{dan84}
several contributions to this evolving new field have been published
with applications to nuclear matter \cite{hsk95,hsk96,hsk95a,hsk95b}, to
one- and
two-band semiconductors \cite{bin97,nai98},to phonon-production in e-e
collisions in plasmas \cite{hsk97} as well as to electron plasmas in
general \cite{bon95,bon97}.
Details of the computational methods is published in Computer Physics
Communications\cite{hsk99}.

The KB-equations are designed to study time-dependent non-equilibrium
phenomena but they can also be used to study the system in its final
equilibrium state. An example is found in  \cite{hsk95} where an
initially un-correlated zero-temperature Fermi-distribution time-develops
into a correlated system whilst the collisions are calculated with
the time-evolving correlated Green's functions.
The build-up of correlations is manifested as an asymptotic decrease
in potential (correlation) energy (initially zero), with a corresponding
increase in kinetic energy, while the total energy is constant.
The resulting Green's functions contain a wealth of information such as
correlation energy and particle distribution. Spectral functions are
also easily derived. Although the collision term basically implies
a second order calculation with respect to the potential the propagators are by the process of
time-iteration dressed with second
order insertions (with their proper energy-dependence) up to all orders.

In this paper we further address and explore these features of the
KB-equations. In particular we focus on correlation times and energies.
We also extend previous comparisons\cite{mor99} with the Levinson equation.
This equation can be obtained from the KB-equations by approximating the propagators
in the collision integral by free Green's functions.
The comparison between the two approaches allow us then not only
to explore the validity of the Levinson equation but also to asses the
importance of the higher order diagrams associated with the dressing of
the propagators. 

Although the interaction can in principle be a
T-matrix, including ladder summations, the presented calculations are
done with an effective time - local interaction.
Section 2 contains a short summary of the KB-formalism used in this
paper. Section 3 deals with the Levinson equation and some relations
involving correlation energies and the Born approximation are shown.
An apparent dilemma concerning the Born approximation result in Section
3 is resolved in Section 4 using the {\it extended quasiparticle picture.}
Numerical results are shown in Section 5 and Section 6 summarizes our
findings.

\section{The KB-equations}
We show some of the equations regarding the KB-formalism needed for our
presentation.  For further details see for example 
refs\cite{kad62,dan84,kra86}. 

In a homogeneous medium neglecting the mean field the KB-equations
reduce (with $\hbar=1$) to:
\begin{eqnarray}
&&(i {\partial\over{\partial t}}-{p^{2}\over{2m}})
G^{^{>}_{<}}({\bf p},t,t') 
=\nonumber\\
&&\int_{t_{0}}^{t} dt''(\Sigma^{>}({\bf p},t,t'')-
\Sigma^{<}({\bf p},t,t''))G^{^{>}_{<}}({\bf p},t'',t')-
\nonumber \\
&&\int_{t_{0}}^{t'} dt''\Sigma^{^{>}_{<}}({\bf p},t,t'')
(G^{>}({\bf p},t'',t')-G^{<}({\bf p},t'',t'))\nonumber\\
&&
\label{eq1}
\end{eqnarray}

\begin{eqnarray}
&&(-i {\partial\over{\partial t'}}-{p^{2}\over{2m}})
G^{^{>}_{<}}({\bf p},t,t') 
=
\nonumber\\
&&\int_{t_{0}}^{t} dt''(G^{>}({\bf p},t,t'')-
G^{<}({\bf p},t,t''))\Sigma^{^{>}_{<}}({\bf p},t'',t')-
\nonumber \\
&&\int_{t_{0}}^{t'} dt''G^{^{>}_{<}}({\bf p},t,t'')
(\Sigma^{>}({\bf p},t'',t')-\Sigma^{<}({\bf p},t'',t')).
\nonumber\\&&
\label{eq1a}
\end{eqnarray}
The notations are the conventional ones. $G^{>}$ and $G^{<}$ are essentially
the occupation-numbers for holes and particles respectively. 
The particle distribution function $\rho({\bf p},t)$ is given by
\begin{equation}
\rho({\bf p},t)=-iG^{<}({\bf p},t,t).
\label{eq2}
\end{equation}
The Green's functions $G^{>}$ and $G^{<}$ are related on the diagonal in
the $t,t'$ plane by
\begin{equation}
G^{>}({\bf p},t,t)=-i+G^{<}({\bf p},t,t).
\label{eq2a}
\end{equation}
Between these two Green's functions there exists a useful relation 
\begin{equation}
G^{^{>}_{<}}({\bf p},t,t')=[G^{^{>}_{<}}({\bf p},t',t)]^{*}.
\label{eq2aa}
\end{equation}
The scattering rates $\Sigma$ are given by
\begin{eqnarray}
&&\Sigma^{^{>}_{<}}({\bf p},t,t')=
-i\int{d^{3}{\bf p_{1}}\over{(2\pi)^{3}}}
G^{^{<}_{>}}({\bf p_{1}},t',t)
 \nonumber  \\
&&\times \left <{1\over{2}}({\bf p-p_{1}})\mid T^{^{>}_{<}}
 ({\bf p+p_{1}},t,t')\mid {1\over{2}}({\bf p-p_{1}})\right  >
 .
 \label{eq3}
 \end{eqnarray}
Here $T^{^{>}_{<}}$ is defined by
\begin{eqnarray}
&&\left <{\bf p}\mid T^{^{>}_{<}}({\bf P},t,t')\mid {\bf p}\right >=
\nonumber \\&&
\int dt''dt'''d{\bf p''}d{\bf p'''}
\left<{\bf p}\mid T^{+}({\bf P},t,t'')
\mid{1\over{2}}({\bf p''-p'''})\right >
\nonumber \\
&&\times
G^{^{>}_{<}}({\bf p''},t'',t''') 
G^{^{>}_{<}}({\bf p'''},t'',t''')
\nonumber \\
&&\times
\left<{1\over{2}}({\bf p''-p'''})
\mid T^{-}({\bf P},t,'''t') \mid {\bf p}\right >.
\label{eq3a}
\end{eqnarray}

% The KB-equations have an equilibrium solution given by \cite{kad62}
%\begin{eqnarray}
%G^{<}({\bf p},t,t')&=&iS({\bf p},t-t')f({{\bf p}^2\over 2 m})
%\nonumber\\
%G^{>}({\bf p},t,t')&=&-iS({\bf p},t-t')(1-f({{\bf p}^2\over 2 m})) 
%\label{equil}
%\end{eqnarray}
%where $S$ is the spectral function defined in  eq (\ref{spectral}) and
%$f({{\bf p}^2\over 2 m})$
%is a Fermi-distribution. The correlations are contained in the spectral
%function while $f$ gives the uncorrelated distribution.

The effective interaction $T^{\pm}$ 
is usually defined in a binary collision (ladder) approximation by
an integral equation formally written as
\begin{equation}
T^{\pm}_{12}=V+VG^{\pm}_{1}G^{\pm}_{2}T^{\pm}_{12}
\label{eq3b}
\end{equation}
where $V$ is the 'free' interaction potential. 
  
In the following $T^{\pm}$ will be approximated by a local
time-independent effective interaction 
\begin{equation}
V({\bf p})=\pi^{3/2}\eta^{3}V_{0}e^{-\frac{1}{4}\eta^{2}p^{2}}
\label{eq3bb}
\end{equation}
with $\eta=0.57 fm$  and $V_{0}=-453MeV$. 
Considering the full dynamical T-matrix approximation one obtains
more involved time integrals which gives rise to nonlocal effects \cite{LSM97,SLM96,MLSCN98}.

\begin{figure}
\centerline{
\psfig{figure=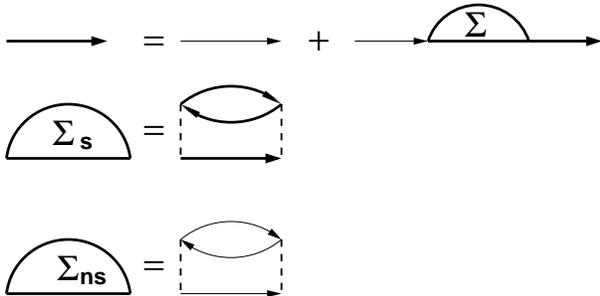,width=8cm,angle=0}
}
%\vspace{.2in}
\caption{
The integral form of Kadanoff-Baym equation (above) and the two
discussed approximations for the selfenergy (below). The first line
describes the selfconsistent Born approximation (thick lines) and the
second line the non-selfconsistent Born approximation (thin
lines). The latter one leads to the Levinson equation (\protect\ref{Lev}).}
\label{sel}
\end{figure}

The exchange term is not included.
The  eq. (\ref{eq3}) for the scattering rates then simplifies to
\begin{eqnarray}
&&\Sigma^{^{>}_{<}}({\bf p},t,t')=
-i\int{{d^{3}{\bf p_{1}}d^{3}{\bf q}}\over{(2\pi)^{6}}}
V^{2}({\bf q})
\nonumber\\&&\times
G^{^{<}_{>}}({\bf p_{1}},t',t)
G^{^{>}_{<}}({\bf q+p_{1}},t,t')
G^{^{>}_{<}}({\bf p-q},t,t').
\label{eq3c} 
\end{eqnarray}
The momentum integrations are conveniently evaluated using the
convolution theorem for Fourier transforms. The diagrammatic
representation can be seen in figure \ref{sel}. We will compare in the
paper the selfconsistent with the non-selfconsistent approximation
which are given in figure \ref{sel} by thick and thin lines respectively.

The total energy of the system reads \cite{kad62}
 \begin{equation}
 E_{\rm tot}(t)={1\over{2}}K_{\rho}(t)+
{1\over{4}}\int{{d^{3}{\bf p}}\over{(2\pi)^{3}}}
( {\partial\over{\partial t_{1}}}
- {\partial\over{\partial t_{2}}})
G^{^{<}}({\bf p},t_{1},t_{2}))
\label{toten2}
\end{equation}
which is used throughout the paper for calculating the total energy.
The kinetic energy $K_{\rho}$ for the correlated medium is
\begin{equation}
K_{\rho}(t)=
\int{{d^{3}{\bf p}}\over{(2\pi)^{3}}}
{p^{2}\over{2m}}\rho({\bf p},t) 
\label{kine}
\end{equation}
and the correlation energy is defined by
\begin{equation}
E_{\rm corr}(t)=E_{\rm tot}(t)-K_{\rho}(t).
\label{corr}
\end{equation}
Note that the mean or Hartree-Fock field is not included in our work and the
total energy  therefore contains only the correlated energy.
We also define the \it uncorrelated \rm kinetic energy by
\begin{equation}
K_{f}(t)=
\int{{d^{3}{\bf p}}\over{(2\pi)^{3}}}
{p^{2}\over{2m}}f({\bf p},t).
\label{kinef}
\end{equation}
The relation between the reduced density matrix $\rho$ and the
quasi-particle distribution $f$ will be discussed below. (See Section 4)
Several numerical applications of this formalism are published. For some
references see the Introduction. A detailed description of numerical
details are published in ref.\cite{hsk99}.

In this paper we concentrate on correlation energies and
correlation times. All (with one exception) calculations are performed with an
initially uncorrelated nuclear matter system and a
momentum-distribution $f({\bf p},t=0)$ specified by a 
density $\rho$ and temperature T.
The system is then time-evolved beyond equilibrium. The selfenergies are
conserving\cite{kad62,bay62} so that the total energy is conserved.
Therefore we have during the time evolution
\begin{equation}
K_{f}^{i}=E_{\rm tot}(t)=E^{\rm eq}_{\rm tot}
\label{kfro}
\end{equation}
where $K_{f}^{i}$ is the kinetic (and total) energy of the 
initial unperturbed and
uncorrelated system 
and $E^{\rm eq}_{\rm tot}$ is the total energy after
equilibration ($t\rightarrow \infty$) including correlations.

\section{The Levinson equation}
As we will show below the Levinson equation can be considered as 
one of several proposed approximations of the
KB-equations. It is of particular interest to us here
because it allows for some analytic results \cite{MLSa97,mor99}
that help to understand the numerical findings
and also help to illustrate some important features built into the KB-equations.
Let us first discuss different reduction schemes of the KB
  equation including the Levinson equation.

To evolve the  two-time KB-equations along $t=t'$ one needs the Green's
functions for off-diagonal times $t\neq t'$.
Different methods have been proposed
approximating the off-diagonal elements in terms of the diagonal
ones and using eq.(\ref{eq2}) a one-time theory can then be derived.

The first method we want to discuss is the Generalized Kadanoff 
Baym (GKB)-ansatz
introduced by Lipavsky et al. \cite{lip86} given by
\begin{eqnarray}
G^{<}({\bf p},t,t')=G^{<}({\bf p},t,t)S({\bf p},t,t')  
\label{eq3c2}
\end{eqnarray}
for $t'>t$ and
\begin{eqnarray}
G^{>}({\bf p},t,t')=G^{>}({\bf p},t',t')S({\bf p},t,t')
\label{eq3c2a}
\end{eqnarray}
for $t>t'$.
$S$ is the spectral function defined by
\begin{equation}
S({\bf p},t,t')=i(G^{>}({\bf p},t,t')-G^{<}({\bf p},t,t')).
\label{spectral}
\end{equation}
The GKB ansatz was discussed in ref \cite{hsk95} and numerical
results were also shown of various approximation schemes.
The very simplest ansatz for the spectral function is the 
quasiparticle approximation to get
\begin{eqnarray}
G^{<}({\bf p},t,t')&=&G^{<}({\bf p},t,t)e^{i\omega(t-t')} \nonumber \\
G^{>}({\bf p},t,t')&=&(-i+G^{<}({\bf p},t',t'))e^{-i\omega(t-t')}.
\label{eq3c3}
\end{eqnarray}
The KB-equations then reduce to  Levinson's equation
\cite{hsk96,L65,L69,JW84} for homogeneous systems
\begin{eqnarray}\label{kinetic}
&&   \frac{\partial}{\partial t} {\rho}({\bf p},t) =
\nonumber\\&&
 2\int \frac{d^{3}{\bf p}_1 d^{3}{\bf p}' d^{3}{\bf p}_1'}{(2 \pi  )^6}
     V(\mid {\bf p} - {\bf p}'\mid )^2
     \delta ({\bf p} + {\bf p}_1 - {\bf p}' - {\bf p}_1')
     \nonumber \\
  &&\times 
  \int_{t_0}^{t }d\tau {\rm cos}
 \Delta \omega (t-\tau)
({\rho}({\bf p}', \tau ) {\rho}({\bf p}_1', \tau )
{\bar {\rho}}({\bf p}, \tau )
{\bar {\rho}}({\bf p}_1, \tau ) 
\nonumber \\
&&
- {\rho}({\bf p}, \tau ) {\rho}({\bf p}_1, \tau ) 
{\bar {\rho}}({\bf p}', \tau )
{\bar {\rho}}({\bf p}_1', \tau ) )
\label{Lev}
\end{eqnarray}
with $\bar {\rho} = 1-{\rho}$ and
 $\Delta \omega=
 (\omega + \omega_{1} - \omega' - \omega_{1}')$ . 
The mean field is neglected in the following and the quasiparticle
energy is approximated by an effective mass $\omega=p^2/2m$.
If for large times the distribution function becomes stationary the 
integration over the cos-function reduces to $\delta(\Delta \omega)$ and
the equilibrium distribution $\rho$ will be a
Fermi-distribution.\cite{mor99}

By the approximation (\ref{eq3c3}) the correlated (damped)
Green's functions in the KB-equations are replaced by the free Green's
functions which in other words means that the second order energy
diagram is calculated without insertions in the
particle (hole) lines.(See Fig. \ref{sel}). 
In the time-evolution of the KB-equations 
the lines are on the other hand dressed repeatedly with $V^{2}$ insertions.
This results in a damping or dephasing while there 
is no damping in the Levinson equation 
but there does remain a memory-effect; the integration over $\tau$ in eq
(\ref{Lev}).

It is instructive to notice that 
the Boltzmann collision integral is obtained from
equation (\ref{kinetic})
if:\\
(i) One neglects the time
retardation in
the distribution functions, i.e. the memory effects
and\\
(ii) The finite initial time $t_0$ is set equal to
$-\infty$ corresponding to what is usually referred to as
the limit of complete
collisions.

For the Markovian Boltzmann equation the kinetic energy is
conserved, while the potential energy is zero. 
The Levinson equation conserves the total energy. \cite{M94}
The correlation energy 
is now given by
\begin{eqnarray}\label{energ}
&&E_{\rm corr}(t) =
\nonumber\\&&
-\frac{1}{4}
\int \frac{d^{3}{\bf p}d^{3}{\bf p}_1 d^{3}{\bf p}' d^{3}{\bf p}_1'}
{(2 \pi  )^{9}}
  V(\mid {\bf p} - {\bf p}'\mid )^2
  \delta ({\bf p} + {\bf p}_1 - {\bf p}' - {\bf p}_1') \nonumber \\
\nonumber\\
 &&\times
\int_{t_0}^{t }d\tau {\rm sin}
                  \Delta \omega
                   (t-\tau)({\rho}({\bf p}', \tau ) {\rho}({\bf p}_1', \tau )
                     {\bar {\rho}}({\bf p}, \tau )
             {\bar {\rho}}({\bf p}_1, \tau ) 
\nonumber\\
&&
- {\rho}({\bf p}, \tau ) {\rho}({\bf p}_1, \tau )
                     {\bar {\rho}}({\bf p}',
                     \tau )
                     {\bar {\rho}}({\bf p}_1', \tau ) ).
\label{ecorr}
\end{eqnarray}
For large times 
eq. (\ref{ecorr}) reduces to \cite{mor99}
\begin{eqnarray}\label{eq}
&&E_{\rm corr}^{\rm eq}=
\nonumber\\&&
-\frac 1 2 \int \frac{d^{3}{\bf p}d^{3}{\bf p}_1 d^{3}{\bf p}' d^{3}{\bf p}_1'}
{(2 \pi  )^{9}}
     V(\mid {\bf p}_1 - {\bf p}_1'\mid )^2
 \delta ({\bf p} + {\bf p}_1 - {\bf p}' - {\bf p}_1') \nonumber \\
&&{\rho}_{eq}({\bf p}_1') {\rho}_{eq}({\bf p}_2')
{\bar {\rho}_{eq}}({\bf p}_1) {\bar {\rho}_{eq}}({\bf p}_2)
  {{\cal P} \over {\Delta \omega}}
\end{eqnarray}
where $\cal P$ denotes the principal value and where $\rho_{eq}$ indicates
the equilibrium large time correlated densities.
This energy resembles the second order Born estimate of the
potential energy but with two important differences. 

The first of these is 
that the densities
$\rho_{eq}$ are correlated densities, in the long time equilibrated limit. 
A Born estimate would however be done with an uncorrelated distribution
$f$. For weak interactions and/or low density for which the Levinson equation
and the Born approximation are certainly valid, the difference between initial
uncorrelated and final correlated densities is negligible. Around
nuclear matter values 
this difference is however important and we 
shall address this question below
showing numerical results for Levinson densities, KB densities 
as well as initially uncorrelated densities.

The second difference between eq. (\ref{eq}) and the second order Born
estimate is that upon closer inspection a factor of one-half appears missing.
This will now be clarified.
With the correlation energy given by eq. (\ref{eq}) the total energy 
after equilibration is using eq. (\ref{corr})
\begin{equation}
K_f^i=E^{\rm eq}_{\rm tot}=K_{\rho}^{\rm eq}+E_{\rm corr}^{\rm eq}
\label{ELevin}
\end{equation}

The second order Born approximation for the total energy is on the other hand
known from perturbation theory 
\begin{equation}
E_{tot}^{eq}=E_{Born}=K_{f}^{eq}+{1\over{2}}E_{\rm corr}^{\rm eq}.
\label{born}
\end{equation}
(As noted above the Hartree-Fock energy is not included so that
accordingly the first order contribution to the energy is not
included here.)
One should
note that in the process of equilibration the system is excited and the
correlation energy does in both expressions (\ref{ELevin}) and (\ref{born})
refer to the excited but not to
the initial ground state of nuclear matter. 
Also note that in the process of excitation the uncorrelated kinetic 
energy $K_{f}^{i}$ has increased to $K_{f}^{eq}$.

In order to resolve the apparent disagreement between eqs. (\ref{ELevin})
and (\ref{born}) we first note that eq. (\ref{ELevin}) results from a
time-evolution of the Levinson equation starting from an uncorrelated
system with a kinetic energy $K_{f}^{i}$. 
We shall show below that the correlated  and uncorrelated kinetic energies
at the end of the time-evolution are related by
\begin{equation}
K_{\rho}^{\rm eq}=K_{f}^{\rm eq}-{1\over{2}}E_{\rm corr}^{\rm eq}
\label{rhof}
\end{equation}
Upon insertion in eq. (\ref{ELevin}) the apparent
disagreement is then resolved.

To prove eq. (\ref{rhof}) we have to discuss the difference between the
reduced density matrix $\rho$ and the quasiparticle distribution $f$.
This is performed in the extended quasiparticle picture.

\section{Extended quasiparticle picture}

In the previous section we  derived the Levinson equation from the 
time diagonal parts of
the Kadanoff and Baym equations. We  adopted the GKB ansatz
for the closure of the off-diagonal parts of the Green's functions. But 
there is another path that we now want to explore.

Our alternative is to use the extended quasiparticle 
approximation (EQP) for the
spectral function in the final equilibrated system. The EQP is
consistent with the Levinson equation as both are a low density and/or weak
interaction approximation but they differ in their physical content
and different renormalizations \cite{MLS00}. 
In an expansion with $Im \Sigma^{+}<<Re \Sigma^{+}$ and
$\partial Re \Sigma^{+}/\partial \omega<<1$
one finds\cite{kra86,hsk92,hsk92a} for the spectral function
 \begin{equation}
  S_{EQP}({\bf p},\omega)=2\pi\delta(\omega-\omega_{0})
   Z({\bf p})
   -{\cal P}{2Im\Sigma^{+}({\bf p},\omega)
   \over{(\omega-\omega_{0})^{2}}}
   \label{eq9.1}
   \end{equation}
   where
   \begin{equation}
    Z({\bf p})=
      1+({\partial Re\Sigma^{+}({\bf p},\omega)\over{\partial \omega}})
      _{\omega=\omega_{0}}
      \label{eq9.1b}
      \end{equation}
is the wave function renormalization 
and where ${\cal P}$ indicates that the principal value is to be taken
when integrating over $S$. The energy $\omega_{0}$ is defined by
\begin{equation}
\omega_{0}=p^{2}/2m+Re\Sigma^{+}({\bf p},\omega_{0})
\label{eq2.1.2b}
\end{equation}
and $\Sigma^{+}$ is the retarded selfenergy.(See also Sect. 5.2). The
EQP approximation satisfies the first two $\omega$ weighted sum rules
\cite{LSM97}.
\begin{eqnarray}
\int {d \omega \over 2 \pi} S_{EQP}({\bf p},\omega)&=&1\nonumber\\
\int {d \omega \over 2 \pi} \omega S_{EQP}({\bf p},\omega)&=& {{\bf p}^2
\over 2 m} + \Sigma_{HF}({\bf p})
  \end{eqnarray}
  and  has been well tested numerically for nuclear
  matter\cite{hsk92}.

In equilibrium the expansion (\ref{eq9.1}) 
is consistent with the following ansatz
\begin{equation}
G^{<}({\bf p},\omega)=if({\bf p})2\pi Z({\bf
p})\delta(\omega-\omega_{0})+\Sigma^{<}({\bf p},\omega){{\cal P}
\over{(\omega-\omega_{0})^{2}}}.
\label{EQP1}
\end{equation}
It allows for an
 approximate construction of the Green's function and the Wigner or
reduced density matrix in terms of the quasiparticle distribution.
The relation between these two distribution functions, the
 quasiparticle distribution and the reduced density matrix,
was  first introduced by Craig
\cite{C66a} within the limit of small scattering rates. An inverse
functional $f[\rho]$ was constructed in ref. \cite{BD68}. For
equilibrium nonideal plasmas the correlated density has been employed in
refs. \cite{kra86,SZ79} and under the name of the {\it generalized
Beth-Uhlenbeck approach} it has been used in ref. \cite{SR87} 
for nuclear matter studies. The authors in refs. \cite{hsk92,hsk92a}
have used this approximation under the name of
{\it Extended Quasiparticle Approximation} for the study 
of the mean removal energy, reduced density and
high-momenta tails of the reduced density matrix. 

The non-equilibrium (time-dependent) extension of this formalism
has been recovered within the 
quasiparticle approach to the kinetic equation for weakly interacting
particles and referred to as a {\it modified Kadanoff and Baym ansatz} \cite{SL94,SL95}.

Integrating eq. (\ref{EQP1}) (in the time-dependent extension) 
over $\omega$ gives a relation between the reduced density
matrix and the quasiparticle distribution
\begin{equation}
\rho({\bf p},t)=f({\bf p},t)Z({\bf p},t)-i\int{d\omega\over{2\pi}}
\Sigma^{<}({\bf p},\omega,t){{\cal P}
\over{(\omega-\omega_{0})^{2}}}.
\label{EQP2}
\end{equation}
The imaginary part of the retarded function $\Sigma^{+}$ is obtained
from
\begin{equation}
 Im\Sigma^{+}({\bf p},\omega,t)=-{i\over{2}}(\Sigma^{>}({\bf
  p},\omega,t)-(\Sigma^{<}({\bf p},\omega,t))
   \label{imsig}
    \end{equation}
    and the real part is obtained from the dispersion relation.
Then
\begin{equation}
 Z({\bf p},t)=1-i\int{d\omega\over{2\pi}}
{{\cal P}
 \over{(\omega-\omega_{0})^{2}}}(\Sigma^{<}({\bf p},t)
-\Sigma^{>}({\bf p},t))
\label{EQP3}
\end{equation}
so that eq. (\ref{EQP2}) can be rewritten as
\begin{eqnarray}
&&\rho({\bf p},t)=f({\bf p},t)-i\int{d\omega\over{2\pi}}{{\cal P}
\over{(\omega-\omega_{0})^{2}}}
\nonumber\\&&
\times
(\Sigma^{<}({\bf p},\omega,t)(1-f({\bf p},t))+
\Sigma^{>}({\bf p},\omega,t)f({\bf p},t))
\label{EQP4}
\end{eqnarray}
providing also the relation between the uncorrelated and correlated energies.
Multiplying with the kinetic energy ${p^{2}\over{2m}}$ and integrating
over ${\bf p}$ one finds the relation (\ref{rhof}) that we wanted to prove.

We like to point out that the first term in eq (\ref{EQP4}) is just the
uncorrelated distribution while the second corrects for the off-shell
scatterings. It was shown in ref \cite{hsk92} that this equation gives
practically exact agreement with a Brueckner calculation. One should
also note that the Levinson equation for the reduced density-matrix
was derived with the ansatz (\ref{eq3c3})
i.e. with a quasi-particle spectral function. The correlations induced
in the time-evolution then result in a spectral function
consistent with the EQP-approximation. If one instead uses the EQP
spectral function in the ansatz the result is
 a nonlocal Boltzmann - like kinetic equation for the
 quasiparticle distribution \cite{SLM96,MLSCN98}. 
For the frequency independent
 real valued interaction used here the nonlocal effects vanish and we
 are left with the Boltzmann equation for the quasiparticle
 distribution
 $f$ from which one then can find the reduced density matrix by
 (\ref{EQP4}). 
 Alternatively one can consider the Levinson equation for the
 reduced density matrix directly. This is the approach taken in this
 paper. For more detailed discussions, derivations and physical content
 of these relations
 we refer the reader to refs \cite{LSM97,MLS00}.

Let us remind that the two outlined schemes, the Levinson equation as
well as EQP, are strictly
valid only at low density and/or for weak interactions. All the same,
the EQP has been found to be an excellent approximation for nuclear
matter\cite{hsk92}.

\section{Numerical results}
In the above we have displayed the KB- as well as the Levinson-equations;
the latter is obtained from KB using free (quasi-particle) Green's functions in the
collision kernel. 

In this section we  show some results of our calculations.
The equations (KB and Levinson) were time-evolved starting at time $t=0$
with an uncorrelated Fermi gas of specified density and temperature.
In addition to KB and Levinson correlation energies we have also calculated
 the second order Born energy given by eq. (\ref{eq}).
In one case we also consider the collision between two Fermi-spheres (slabs in
coordinate-space).

A comparison with the approximate Levinson and Born results requires a
high precision of the calculations to be
meaningful. To minimize the relative errors all calculations are made
with essentially the same KB computer-program. 
To perform the Levinson calculations all that has to be changed relative
to the KB is to
replace the selfconsistently calculated Green's functions used in the
KB-code in the collision-kernel with the free Green's functions 
shown in eq (\ref{eq3c3}).
In the second order Born-approximation calculations the Green's 
functions on the
time-diagonal were  replaced by  constant (time-independent)
distribution functions  $\rho({\bf p})$ as specified in the respective
calculations below.

It may be noteworthy to
point out that the Pauli-blocking which in most perturbative
calculations such as Brueckner is treated approximately (e.g. with an angle
averaging) here is treated "exactly".

The meshes that we use are an improvement relative to previous work as more
computer-power is now available.  The momentum-mesh in the Cartesian
coordinate system that we use was 57 points along each  axis with
$\Delta p_{x}=\Delta p_{y}=\Delta p_{z}=0.2 fm^{-1}$. 
The number of time-steps was typically $80$ with
$\Delta t=0.25fm/c$ .
The interaction shown in eq. (\ref{eq3bb}) was used. 
The dependence of the strength $V_{0}$ of the interaction is now also
investigated.

\subsection{Approach to Equilibrium; Correlation Times}
If one time-evolves an initially uncorrelated Fermi-distribution
with the classical Boltzmann-equation (modified for Fermion-statistics)
this initial distribution would be stationary. The Fermi-distribution
is in fact a solution of this equation with the collision-term equal to
zero. There are no two-body correlations. This is not so for 
the Kadanoff-Baym equation. 
As the system in this case evolves from the initial uncorrelated state the 
correlation energy decreases with time.  The system correlates.
At the same time the kinetic energy  increases by the same amount,
because our choice of self-energies is conserving i.e. the total energy is 
conserved. As a
consequence of the self-consistent time-iterations higher and
higher orders of insertion diagrams are included in the Green's function
propagators until full selfconsistency and equilibration is achieved.
%The equilibrated Green's functions satisfy eq. (\ref{equil}).
In the KB-case the system converges to an equilibrated distribution but
this is not always so for the Levinson-case. We shall return to this
below.

In a previous publication\cite{mor99} we have already considered the correlation
time $t_c$, i.e. the time it takes the system to correlate
time-evolving from the initially uncorrelated Fermi-distribution.
It was in particular found that in the Levinson-approximation a simple
result could be derived at low temperature namely
\begin{equation}
t_c={1\over{E_F}}
\label{tc1}
\end{equation}
with $E_F$ the Fermi-energy. The temperature-dependence was found to be
relatively small.
This result was also used in a subsequent paper on interferometry
methods\cite{mor00}. 

\begin{figure}
\centerline{
\psfig{figure=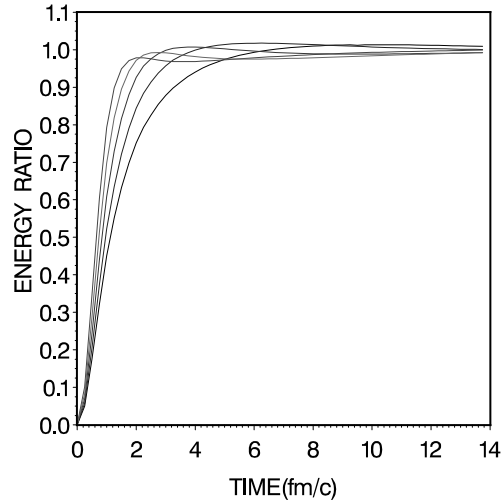,height=10cm,angle=0}
}
%\vspace{.2in}
\caption{\label{1}
Correlation energy from KB-calculations as a function of time 
normalized to final
equilibration energies. From left to right the five curves correspond 
to the normal density  of nuclear matter ($\rho_{0}=0.18 fm^{-3}$)
multiplied by
$2,1,0.5,0.25$ and $0.125$. See also Table \ref{tab1}}
\end{figure}

Fig. \ref{1} shows  the correlation energies normalized to the
equilibrium energy as 
a function of time for five different
densities as indicated in the figure-caption.  The initial distribution
is in each case a zero temperature Fermi-distribution. 
In accordance with ref. \cite{mor99} we define the correlation times
$t_c$ as the time of maximum correlation-energy.
We now find that the  times $t_c$ scale roughly as 
\begin{equation}
t_c={1\over{2E_F}}
\label{tc2}
\end{equation}
as shown in Table \ref{tab1}. The previous analytic result (\ref{tc1})
based on the
Levinson equation gave correlation times twice as large.
A precise definition of these times is however not possible
as is also seen in figure \ref{1}.

\begin{figure}
\centerline{
\psfig{figure=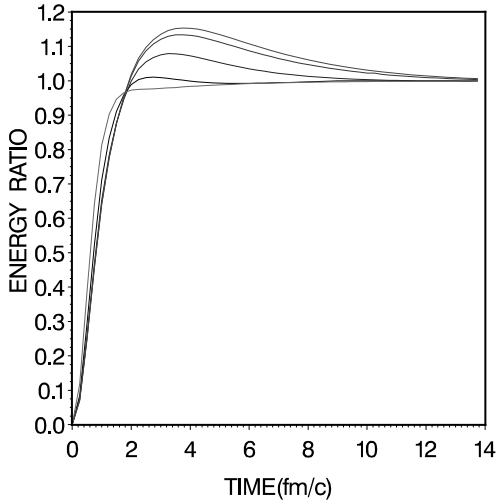,height=10cm,angle=0}
}
%\vspace{.2in}
\caption{\label{2}
Similar to Fig. \ref{1} but for normal nuclear matter density and
different interaction strengths. The normal strength $V_0=453$ is multiplied by
$2.0,1.0,0.5.0.25$ and $0.125$ respectively.  The figure shows  t
there is an increasing "overshoot" of the correlation energy as the
strength is decreased. See also Table \ref{tab3}}
\end{figure}

Fig. \ref{2} shows  the correlation energies normalized to the
equilibrated energy at five different interaction-strengths
indicated in the figure-caption.  The initial distribution is in
each case a zero-temperature Fermi-distribution.
The previous analytic finding \cite{mor99}was that the correlation
times are roughly independent of the interaction strengths. 
This is nicely confirmed by the numerical results of Fig. \ref{2} 
  for times smaller than $2$~fm/c.
There is
however a large overshoot as the strength is decreased. This still remains
to be understood. 
With the particular definition of the correlation-times 
that we have adopted we still
see a dependence on the interaction-strengths because of this overshoot
as displayed in Table \ref{tab3}.

\begin{figure}
\centerline{
\psfig{figure=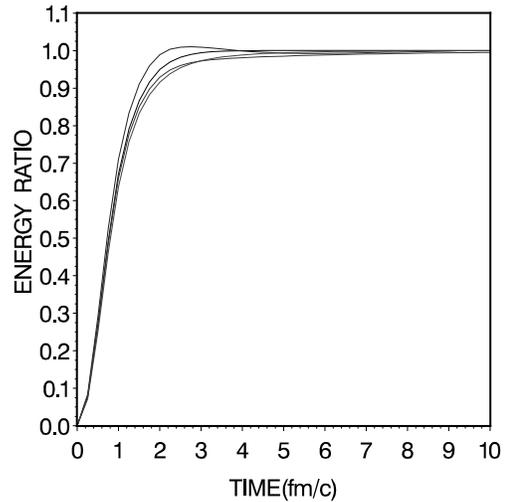,height=10cm,angle=0}
}
%\vspace{.2in}
\caption{\label{3}
Correlation energy as a function of time normalized to final
equilibration energies.
The interaction strength and density are both normal. From left to right
the first,third and fourth curves are for initial temperatures
  $ 0,20$ and $40$  MeV respectively. The second curve is for a
    "collision" event (see text). Notice the near overlap of the
    curves.}
    \end{figure}

In a previous paper it was also shown that the
temperature dependence of $t_c$ should be  quite small. \cite{mor99}
This conclusion was based on the Levinson equation.
The result using the KB-equations 
shown in Fig. \ref{3} confirms this. 
One may only note that the approach to
equilibration is markedly different at higher temperatures
as compared to zero temperature. 
Our general conclusion regarding the independence of temperature 
is also exemplified by the second curve from the left in Fig.
\ref{3}. This shows the correlation energy in a "collision" between two
Fermi-spheres separated by a momentum of $2.2 fm^{-1}$ (100 MeV/A
collision energy) and with a total
density of $0.18 fm^{-3}$ (normal density).

In the calculations with the KB-equations 
our choice of selfenergies are conserving, i.e.
total energy is conserved. \cite{bay62} 
The Levinson-equations also conserve total energy. From our 
calculations we find after $80$ time-steps
a decrease in total energy of $3.8\%$
at normal nuclear matter density
 and similarly an  increase of $1.3\%$ in the KB-case.
This numerical accuracy is quite satisfactory.

\begin{figure}
\centerline{
\psfig{figure=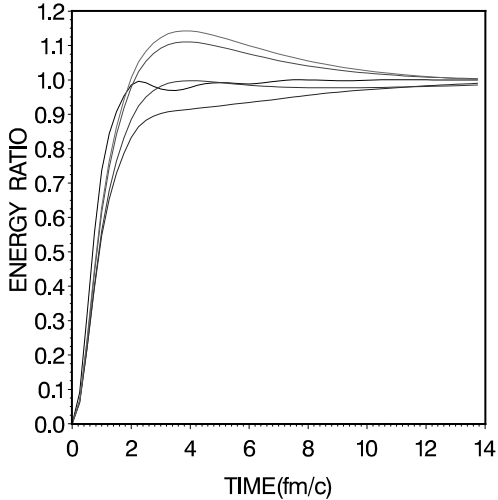,height=10cm,angle=0}
}
%\epsfxsize=8cm
%\vspace{.2in}
\caption{\label{4}Time-dependence of normalized correlation energies in
the
Levinson-case for normal nuclear matter density. The oscillating curve
is with the interaction strength twice the normal. The smooth curves
from bottom and up are with strengths
 $1.,0.5,0.25$ and $0.125$ times the normal. Notice that  the 
 correlation energy converges only for the two weakest interactions.
 Compare with the KB-results in Fig. \ref{2}.  The initial distribution
 is in each case a zero-temperature distribution of normal nuclear
 matter density.}
\end{figure}

Another important result shown
by Figs 2-4 is that the correlation energy (and the kinetic energy)
converges to an equilibrium value in the KB-case.
While the KB-results show excellent convergence this is not true in the
Levinson case as shown by Fig. \ref {4} which should be compared with
Fig. \ref{2}. One finds a convergence only for
the two weakest interactions  which also agree quite well with the
corresponding curves in Fig. \ref{2} while only moderate
convergence with the factor $0.5$. For the strongest interaction
there is even sign of an instability in agreement with the findings of
Haug \cite{BTRH92} and others \cite{KBKS97},\cite{P98}
who found a continuous increase of energy with time.  The origin of
this problem with the Levinson equation was discussed in terms of the
artificial collisional broadening \cite{SPLI95}. 

We note that in the Levinson case the Green's functions are
free, uncorrelated, in the collision-term  while the ensuing 
Green's functions are not free as
e.g. evidenced by the non-zero potential energies. This is an
inconsistency in the Levinson equation which, we believe, is related to
the non-convergence. 
 The problem with this equation is
also reflected in
 the results
 in Table~\ref{tab3}
 showing reasonably good agreement between KB and Levinson correlation
 energies only for the three weakest interactions. 

The fact that the correlation time is finite and of the order of some $fm/c$
in nuclear matter is, we believe, of importance when studying heavy ion
collisions. Correlations will change with density and temperature with
typically this response time. In a previous paper we have already looked at
the consequence of this fact on interferometry measurements.\cite{mor00}

\subsection{Equilibrium Spectral Function; Occupation Numbers}

Correlation energies are intimately related to the spectral functions in
our case given by eq. (\ref{spectral}).
In the quasi-particle approximation valid for weak interactions and an
uncorrelated medium they are simply delta-functions corresponding to
vanishing widths represented by the first part of eq. (\ref{eq9.1}). 

In this section we consider the system in its final equilibrated state.
The total energy is then related to the spectral function  by
\begin{eqnarray}
 E_{\rm tot}&=&2\int_{-\infty}^{+\infty} d\omega f(\omega) \int_{-\infty}
 ^{+\infty}\frac{d^{3}{\bf p}}{(2\pi)^{3}}
 [{p^{2}\over 2m}+\omega]S({\bf p},\omega) 
 \label{toten1}
 \end{eqnarray}
which is the same as
eq. (\ref{toten2}) with the equilibrium relations
 \begin{eqnarray}
 G^{<}({\bf p},\omega)&=&iS({\bf p},\omega)
 f(\omega)\nonumber \\
 G^{>}({\bf p},\omega)&=&iS({\bf p},\omega)(1-f(\omega)).
 \label{distr}
 \end{eqnarray}
% The last equations are simply eqs (\ref{equil}) but in $\omega$-space.
A factor $4$ is included for the spin, isospin etc. degeneracy of 
the nuclear system.

\begin{figure}[h]
\centerline{
\psfig{figure=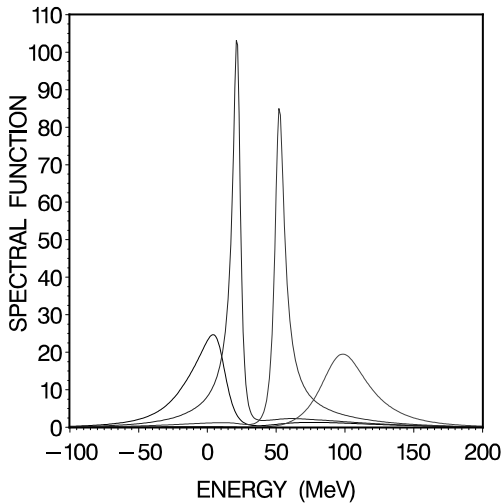,height=10cm,angle=0}
}
%\vspace{.2in}
\caption{\label{6}
Similar to Fig \ref{5} except that these spectral functions are for the
ground-state and obtained with imaginary time-stepping.
The momenta are from left to right $.0,1.0,1.6$ and $2.2 fm^{-1}$. }
\end{figure}

\begin{figure}[h]
\centerline{
\psfig{figure=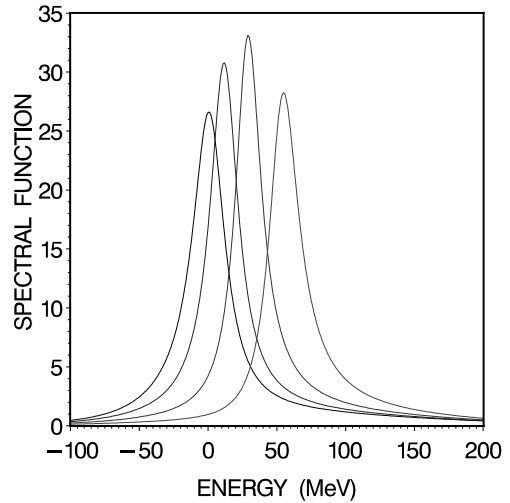,height=10cm,angle=0}
}
%\vspace{.2in}
\caption{\label{5}
Spectral functions  in nuclear matter at $\rho=0.18 fm^{-3}$, $k_{f}=1.4
fm^{-1}$. The
momenta are from left to right $0.0,0.75,1.2$ and $1.65 fm^{-1}$. Note
that the last momentum is larger than $k_{f}$. The temperature is here
estimated to be $25$ ~MeV. }
\end{figure}

The spectral functions $S({\bf p},\omega)$ 
can be calculated from eq. (\ref{spectral})  by
Fourier-transforming from the time-domain. It was found more practical
however to use
\begin{equation}
\ S({\bf p},\omega)={2 Im\Sigma^{+}({\bf p},\omega)\over{(\omega
  -{p^{2}\over{2m}}-Re\Sigma^{+}({\bf p},\omega))^{2}+
    (Im\Sigma^{+}({\bf p},\omega))^{2}}}.
    \label{eq2.4.2}
    \end{equation}
Figure \ref{5} shows some results at normal
    nuclear matter density. These are from KB-calculations. 
    (As noted above in section 3 the spectral functions in the Levinson
    approximation are in agreement with the EQP-approximation
    (\ref{eq9.1}) in lowest order densities.) The initial distribution 
    is a zero-temperature
    Fermi-distribution and the $\omega$-dependent selfenergies are 
    obtained by Fourier-transforming from the time-domain after
    equilibration. The correlation energy is in this case {35.95 MeV} as
    shown in Table~\ref{tab1}.
The figure shows that the widths are comparable with those from
    Brueckner and other many body calculations with "realistic"
    interactions. 
    If the calculation
    were for the ground-state the distribution would be much more peaked
at this momentum. Because the calculations in our present work are
made from an initially uncorrelated state that is then time-evolved along
the real axis the final state corresponds to an excited state  (with a
temperature estimated to be about 25 MeV).  For
comparison we also show in fig. \ref{6} 
spectral functions for the ground state obtained with 
imaginary time-stepping.\cite{dan84,hsk95}
The momenta at the Fermi-surface is not shown here as it would be too high and
narrow in this case, but note the peaks at the adjacent momenta.

The occupation-numbers  are in equilibrium also related to the
spectral functions by
\begin{eqnarray}
 \rho({\bf p})&=&{1\over{2\pi}}
  \int_{-\infty}^{+\infty}f(\omega)S
  ({\bf p},\omega)d\omega 
  \label{occup1}
  \end{eqnarray}
consistent with eqs. (\ref{eq2}) and (\ref{distr}).

\begin{figure}
\centerline{
\psfig{figure=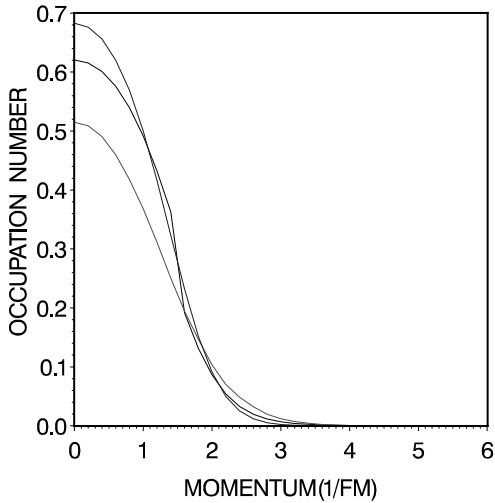,height=10cm,angle=0}
}
%\vspace{.2in}
\caption{\label{7}
Density distributions from the KB-  and Levinson- 
calculations at normal nuclear matter density. The KB-distribution is
larger at small momenta and smaller at larger momenta. Notice that it
shows the characteristic discontinuity at the Fermi-surface
that is related to the strength-function. Compare with Fig. 1 in ref.(3)  .}
The Levinson curve is smooth (it is a Fermi-distribution)
and starts at about $0.52$ at p=0. For
comparison is also plotted an estimate of the  Fermi-distribution $f$ 
with the same nuclear matter
density and a temperature of $27$ MeV. It starts at about $0.68$.
\end{figure}

 Figure \ref{7} shows a comparison of the density-distributions obtained
respectively from the KB- and Levinson calculations. Please note that
the KB calculations lead to a pronounced
discontinuity at the Fermi energy. A quasiparticle (Fermi-)
distribution $f$ with a temperature of $27$ MeV is plotted for
comparison.  This is the roughly estimated temperature of the final
equilibrated system. This would be the distribution in a Boltzmann
calculation in which case there are no correlations, the spectral
function is the quasi-classical and the $\rho$ and $f$ distributions are
identical.
The KB - as well as
Levinson distributions possess power tails at high
momenta and suppress states at lower momenta due to
correlations\cite{LSM97}.

Correlation energies are shown in Tables \ref{tab1} and \ref{tab3}.
These are energies  after $80$ time-steps ($20 fm/c$).
At this time the system is well equilibrated in the case of KB
 as  already seen in the figures \ref{1}, \ref{2} and
\ref{3}.
In other words the Green's functions along the time-diagonal become
constant for large
times. 

\begin{figure}
\centerline{
\psfig{figure=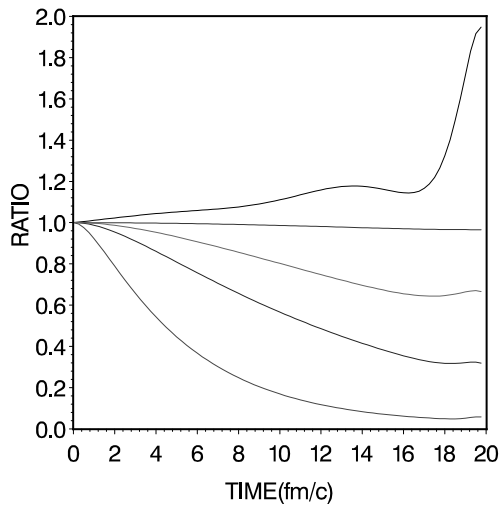,height=10cm,angle=0}
}
%\vspace{.2in}
\caption{\label{8}This figure shows the absolute value of 
$G^{<}({\bf p},T,t')/G{<}({\bf p},T,T)$ for $T-t'$ for four 
different strengths of
the interaction. The four lower curves with the ratio less than 1.0 is
from KB-calculations. From bottom to top the normal strength is multiplied by
factors of $2.,1.,0.5$ and $0.125$. The uppermost curve is from Levinson
calculation with normal interaction strength  showing no damping but
rather an increase with past times. This effect is explained in the
text. The momentum is chosen to be ${\bf p}=0$ for all cases.}
\end{figure}

This is not so for the off-diagonal elements that carry the
correlations. Fig
\ref{8} shows the absolute value of these elements normalized to the equilibrium
value, as a function of past times.   For the KB-calculations the
well-known damping or dephasing is seen.
The slight increase for the largest past times is a natural
consequence of the correlation process. While the equilibrated
occupation for $p=0$ is  $0.62$. (See fig \ref{7}) the initial value was
$1.0$ and this is still "memorized" $20 ~fm/c$ later. 

Of particular interest is the
increase in damping as the strength of the interaction is increased
which is well illustrated by the figure.
This is a very important
effect that is contained in the KB-equations with the selfconsistently
calculated Green's functions.  
This has as a consequence that the memory-time decreases;
the integration over past times which in
principle should start at the time when interactions are switched on,
($t_{0}$ in eqs. (\ref{eq1} and \ref{eq1a})), can 
now be limited to about $5 fm/c$ or less 
before the point of time of the evolution.
% if the interaction is of normal strength. 
Fig 9 in ref.\cite{hsk95} 
illustrates this point. Associated with an increase of the strength of
the interaction is of course not only a decrease in memory-time but also
an increase in the width of the spectral functions. 

Only for the
very weakest strength  the damping is negligible. This
is consistent with Table \ref{tab3} that shows a good agreement between 
all methods for this case. If the situation were such that 
the evolution at the same time is sufficiently slow another 
limit is reached. The
time-integration can now be done analytically the energy-spread
decreases and collapses to a delta-function. The Boltzmann limit is
reached.

The damping, so characteristic in the KB-case is absent in the Levinson
case.
The corresponding curves can increase or decrease in the latter case
depending on the past
distribution-function (on the time-diagonal).  In the present
calculations  this
function will always decrease for ${\bf p}=0$ and the plotted ratio in
Fig. \ref{8} is now consequently larger than 1.0. 

\subsection{Equilibrium correlation energies}

It was already shown above that the Levinson correlation energy for
large times approaches a second order Born value.
But it was also pointed out  with reference to eq. (\ref{eq}) that
the Born calculation then has to be made with the Levinson final
reduced density but {\it not} with the initial
quasiparticle distribution as would normally be done in a perturbative
expansion. In the low-density and/or weak interaction
limit, where the Born and
Levinson approximations both become valid this difference should be
irrelevant. This is verified by the results shown in columns 
5,6 and 7 in table \ref{tab1} showing Born energies
calculated with occupation numbers from the KB, Levinson and initial
distributions  (labelled "KB","Lev." and "Init.") respectively. 
As expected the three energies agree
exactly at the lowest density but the agreement becomes progressively
worse as the density is increased.  As predicted by eq.
(\ref{eq})
the Born-column indicated "Lev." (column 6)
agrees  on the other hand nearly exactly with the Levinson result 
(column 4) at all densities, being a confirmation of computational accuracy. 

\subsubsection{Comparison of Levinson with Born}

As noted above the Born(Lev) correlation energies agree very well with the 
Levinson energies at all densities and temperatures but they differ from the
Born(Init) energies as seen comparing columns 6 and 7 in table \ref{tab1}.
This is because the occupation-numbers differ in these two calculations.
The Born(Init) calculations are made with the Fermi-distributions of the
temperatures T indicated. These are also the {\it initial} distributions
in the Levinson calculations, but the final distributions $\rho$ (with
which also Born(Lev) is calculated) are different. 
There are two distinct effects contributing to the change from
initial to final distribution.
One is the heating associated with the correlations and the 
second comes from the
difference between the reduced density matrix $\rho$ and the quasiparticle
distribution $f$ that is contained in the off-shell part in EQP (\ref{EQP4}).
We are going to discuss these two effects now.

The heating of the system in the Levinson case occurs since it correlates. 
The uncorrelated quasiparticle distribution $f$
is consequently of a higher temperature in the Born(Lev) case, column 6, 
than in the Born(Init) case, column 7.

In order to make a meaningful comparison between the two
results we have to correct for this temperature-difference. Fortunately
this is straightforward in the Levinson case.
 The numerical solution of the
Levinson equation gives us the correlated distribution 
$\rho({\bf p},t) $ and from eq. (\ref{kine}) the kinetic energy 
$K_{\rho}(t)$ in the equilibrated medium (for large times).
 The relation (\ref{rhof})  
valid for the Levinson equation then gives us the kinetic energy
$K_{f}^{\rm eq}$ which together with the density allows us to deduce the
relevant parameters, temperature and chemical potential for the {\it
final} equilibrium Fermi-distribution. 

The Levinson correlation energy is then compared with the 
Born(Init) correlation energy calculated with that same final
Fermi-distribution.
The result of this comparison is shown in Table \ref{tab2} at normal
density.
With the Born and Levinson calculations compared at the same temperature
($T_{f}$ in the table) 
i.e. with the same distribution $f(\omega)$ there is however a remaining
difference, in Table \ref{tab2} indicated by $Diff.$ 
We remind that when Born
is calculated with the $\rho$-distribution (i.e. in Born(Lev)) rather than
with the $f$-distribution we get no difference, i.e the Levinson energy. 
The remaining difference $Diff$  is then due to the
second of the two separate effects referred to above. It is
attributed to the difference between the reduced density $\rho$ and the
corresponding quasiparticle distribution $f$. 
The  former contains the spectral correction as in eqs. (\ref{EQP2}) and
(\ref{occup1}).

Such a spectral correction can be included in e.g. Brueckner type
calculations by iteration but is rarely done.  (See however a recent
work \cite{Ro00}).
Table \ref{tab2} shows 
a decrease of $3.3 MeV$ of the correlation energy at zero temperature.
This implies a decrease in binding
energy of $1.6 MeV$. 
(We like to point out that in the EQP-approximation a change 
$\delta$ in correlation energy for a given
distribution $f({\omega})$ changes the total energy by $\delta/2$. It
changes the kinetic energy $K_{\rho}$ by $-\delta/2$.)

The dependence on the strength of the interaction is presented in 
Table \ref{tab3}.
The Born calculations are here for 
initial zero temperature Fermi distributions.
For the three weakest interactions all three correlation energies agree
reasonably well. 
But at full interaction strength we see differences of over 20 \% 
increasing to 50 \% at twice the strength.

\subsubsection{Comparison of KB with Levinson}

We now want to compare the KB- with the Levinson- equation.
Although the KB collision term in our calculations  is formally up to
second order in the interaction the correlation energy at equilibrium 
is actually of a much higher (infinite) order. This is  because the correlated
Green's functions are formed by
iterative time-stepping functionals of the interaction.  

The effect of these higher order terms can be assessed from comparing
the KB and Levinson correlation-energies. The
difference between the two stems from the difference between the Green's
functions in the collision kernel for the two separate cases. 
In the KB case they are
selfconsistent (correlated) while in the latter they are free Green's
functions. In diagrammatic language the presence of the
correlated Green's functions in the collision kernel means that  hole- 
 and  particle-lines are
dressed  with the two-point (second order) insertions {\it to all orders}
and {\it all time orderings}, see Fig. \ref{sel}.  
This implies that for a similar
calculation in the
$\omega$-representation the
proper dependence on $\omega $ of the selfenergy 
$\Sigma^{+}({\bf p},\omega)$ has to be included in the calculation. The
same problem is faced if applying Brueckner's theory in his original
recipe where the insertions in particle lines are calculated off the
energy-shell.\cite{bru55,bru58,co65,hsk73}
This is quite complicated and is a compelling reason 
why the 'continuous' recipe (with both particle- and hole-insertions
being on the energy-shell) is now customarily used. It seems that
the numerical problem with off-energy shell propagation is simplified by
going to time-space, as in the present work.

 In Table \ref{tab1} one finds that the difference between the "KB"
and "Lev" correlation energies in columns 3 and 4 at normal density
amounts to $13.72~ MeV$ for an initial temperature
$T=0$ and decreases to $5.63~ MeV$ at an initial $T=60 MeV$. This implies
a difference in binding energy of $6.86$ and $2.82$ MeV respectively.
(See discussion above regarding  factor of $1/2$.)
One also finds in the same table that the
difference decreases with density which is to be expected. The
difference in binding energy would be half
of the quoted numbers. 
The conclusion is
that the higher order diagrams resulting from insertions in the
propagators contained in the KB eqs but neglected
in Levinson are quite
important.

\subsection{Discussion}

Referring to the discussion above in table \ref{tab2}, our results 
do actually consider excited, not ground state 
nuclear matter. In the Levinson case we could easily estimate the
excitation temperature.(See section {\bf V.B}). 
In the KB-case there is no such simple relation. Using the Levinson
relation as a rough estimate we find however an excitation energy of
about $25$ MeV at normal density. 
Ground state
calculations can be done by imaginary time-stepping but this has not yet
been done. The calculation showing the zero temperature spectral
functions in Fig. \ref{6} is not 
sufficiently accurate to allow a detailed comparison of correlation energies. 

We do expect the relatively large difference between the Levinson and KB
correlations  to prevail at zero temperature. 
We also argue that our Levinson calculation closely corresponds to a 
'continuous' choice (on the energy-shell mean field insertions in hole- and
particle-lines) Brueckner calculation.
With this choice
the higher order Brueckner diagrams that usually are considered (three-body,
low-order ring and some fourth order) then contribute only a 
few MeV.\cite{So97}  
The  Brueckner mean field insertions mentioned above are 
of course not present in 
the Levinson case but because we use a local interaction they are
expected to give a negligible correction.
We have confirmed this numerically. With
realistic non-local (momentum-dependent) effective interactions one has
of course
important dispersion-corrections.

With the self-consistent Green's functions as in the KB-calculations the
width of the spectral functions is (implicitly) included. In contrast
the Brueckner calculations are done in a quasi-classical
approximation with the effective interaction, the reaction-matrix,
calculated assuming an uncorrelated zero-temperature Fermi distribution.
Using Green's function methods normally including also hole-hole
ladders one also rarely goes beyond the first quasi-classical approximation.
(Except for the hole-hole diagrams this is equivalent then to Brueckner.)
Spectral-functions are however readily obtained and the equations can be
iterated. This is a rather formidable calculation, but has recently been
done with realistic forces.\cite{Ro00}
The KB-calculations in this paper corresponds to such an
iterated calculation with selconsistent spectral-functions and off-energy
shell $\Sigma$'s albeit with a simple interaction and no ladder
summation. The results from the
two calculations are not readily compared but both agree in that the
corrections going beyond the quasi-classical approximation should be
considered.

Referring to Fig. \ref{8} an important difference between the Levinson
and KB-collision terms is the damping, which is related
to the width of the spectral-function. Referring to eq. (\ref{eq2.4.2})
this is on the other hand related to $\Sigma^{+}({\bf p},\omega)$. 
Above we deduced a $1.6~ MeV$ difference in binding energy
between the Born (Init)
and the Born (Levinson) calculations stemming from a spectral correction
of the time diagonal density-matrix. The present difference between
Levinson and KB of $6.86~MeV$ is also a
spectral correction but stems from the off-diagonal elements i.e. the $\omega$
dependence of $\Sigma^{+}$.

\section{Summary and Outlook}
The Kadanoff-Baym equations have been used 
to calculate correlations and to find the effect of approximations such as
using undressed propagators.
The KB-calculations as presented above are fairly simple.
The fortran version of the program is available from Computer Physics
Communications\cite{hsk99}.

In our KB- as well as Levinson-calculations the initial conditions are
in general
Fermi-distributions of specified temperature and density. As the
equations
are time-stepped the system correlates with a decrease in the potential
energy. The total energy is conserved and the kinetic energy increases.
The temperature
of the system is consequently increasing. The final correlation-energies
are therefore not for the ground-states of the respective systems.
To explicitly study the ground-state one can use the imaginary
time-stepping
technique.\cite{dan84,hsk95} It would have been desirable to use this in
our present work for calculation of correlation energies.
The precision is however not (yet) at the level of the
computations with the uncorrelated initial condition. Here we only used
the imaginary time-stepping method to calculate
the spectral function at zero temperature shown in Fig. \ref{6}.

One of the purpose of this work was to investigate the time
it takes for the system to correlate from the initially uncorrelated
state. We have in particular calculated the dependence upon density,
temperature and also strengths of the interaction as shown in Sect.
5.2. We find that the correlation-time $t_{c}$ scales roughly as
$$t_c={1\over{2E_F}}$$
but is nearly independent of the strength of the interaction and of
the temperature of the system. This time is relevant for the discussion
of collisions between heavy ions when $t_{c}$ is comparable with the
collision-time between the ions.
It is also of practical importance in dynamic calculations. In principle
the calculation with the KB-equations involves an integration over all
past
times referred to as a memory-effect. In practice this is not necessary.
The correlations effectively cuts down the memory-time. This is also
demonstrated by the damping shown in Fig. \ref{8}. In nuclear systems
the memory-time is typically less than $5 fm/c$, see ref.
\cite{hsk95}, or 10-20
time-steps. In electron plasmas where correlations are smaller the
corresponding time is $100 fs$ or  25-50 time-steps.

It was shown in a previous work that the Levinson correlations approach
a second order Born expression at large times.\cite{mor99} Increasing
the strength of the interaction an instability does however develop as
shown in Fig. \ref{4}. A dilemma involving a factor of $2$ comparing
with the Born expression for the correlation energy was resolved in Sect
4. In order to compare the Levinson with the Born correlation energies a
temperature correction had to be applied and the result was shown in Table
\ref{tab2}. There remains a difference between the two which amounts 
to a decrease in binding of $1.6 MeV$ This can be
ascribed to a spectral correction and is due the difference
between
the uncorrelated distribution $f({\bf p})$  and the
reduced density $\rho({\bf p})$  in the Levinson calculation. The
relation between $f$ and $\rho$ is expressed by eq. (\ref{occup1}) which
in the EQP approximation reduces to eq. (\ref{EQP4}).

A comparison of the KB- with the Levinson-correlations show the effect
of the dressing of the propagators by the $V^{2}$ insertions. This is
found to result in a comparatively large correction; a $6.85~MeV$
decrease in binding energy. This correction is
also ascribed to a spectral correction but now for time off-diagonal
elements. This result should have an impact on the ongoing problem of
nuclear saturation which is found to be above the experimental density
and/or energy for perturbative schemes without spectral corrections.

The total energy given by eq. (\ref{toten2}) includes besides a kinetic energy
only the binary correlation energy. 
There is of course also a first order term, the Hartree-Fock term,
contributing to the energy of the many-body system.
This  mean field term can be included easily in the
Green's functions.\cite{hsk99} 
We point out that besides the usual momentum independent Hartree shift
another effect appears usually referred to
as the dispersion-effect. This effect stemming from the
momentum-dependence of the Fock (or Brueckner-Hartree-Fock)
field is well-known and not of interest in our present calculation. 
It further decreases the binding energy in nuclear many-body calculations.

All of our nuclear matter calculations have  until now been restricted
to using a time-local interaction. The correlations
appear however to be similar to those for more realistic interactions.
This is illustrated by the spectral functions shown in Figs. \ref{5} and
\ref{6}. They show a width comparable to more serious calculations and
the expected behavior as a function of momentum.\cite{hsk92}
Our interaction does not have a short-ranged repulsion or tensor-part
but we believe the long-ranged part to be a reasonable representation of
the true interaction. We do however
envision a future extension to more realistic nucleon forces as well as
a
T-matrix.

It may finally be relevant to point out that an important difference
between the present  KB-calculations
and more conventional Green's function studies is that it has been
performed here in $t,t'$
rather than $\omega$ -space. This is found to be very practical.

This work was supported in part by the National Science Foundation Grant
No. PHY-9722050

\newpage
\onecolumn
\begin{table}
%\squeezetable
\caption{\label{tab1}}Correlation energies as a function of the density
of nuclear matter. At normal density the temperature dependence is also
shown. All results are here with $V_{0}=453.0 {\rm MeV}$.
The energies $E_{\rm corr}^{\rm eq}$ are (the negative of) the
equilibrium correlation energies.
The Born energy $E_{\rm corr}^{\rm eq}$ (Born) is calculated with three
different distribution-functions as discussed in the text. 

\begin{tabular}[t]{|c|c|c|c|c|c|c|c|c|c|}
\hline
& $\rho$ &  $ T$  & $E_{\rm corr}^{\rm eq}$  (KB)  &
$E_{\rm corr}^{\rm eq}$ (Lev)& \multicolumn{3}{c|}
{$E_{\rm corr}^{\rm eq}$ (Born)}&
$t_c$ (KB) &${{\hbar\over{2E_{F}}}}$
\\
&${\rm fm}^{-3}$&MeV& MeV& MeV
&\multicolumn{3}{c|}{MeV}&${\rm fm/c}$&${\rm fm/c}$
\\
&&&&& KB& Lev.& Init.&&
\\ \hline
 &0.380&0&53.63& ---&-&-&65.16&2.0&1.5
  \\ \hline
   &0.183&0&35.95&49.67&-&49.69&43.97&2.4&2.4
   \\ \hline
     &0.181&10&36.03&48.60&50.52&48.54&49.16&---&---
        \\ \hline
            &0.182&20&35.94&46.74&--&46.65&52.84&---&---
            \\ \hline
              &0.182&40&34.31&42.14&--& 42.09& 50.52&---&---
                  \\ \hline
                      &0.182&60&31.59&37.22&38.00& 37.20& 44.65&---&---
                      \\ \hline
                        &0.095&0&23.55&31.33&--& 31.76& 28.83&3.4&3.8
                           \\ \hline
                                &0.047&0&14.40&17.84&18.24& 18.26&
                                17.47&5.2&6.2
                                \\ \hline
                                  &0.023&0&8.42& 9.80&9.96& 9.96&
                                  9.96&8.5&9.7
                                    \\ \hline
                                              \end{tabular}
                                                        \end{table}

\begin{table}
\caption{\label{tab2}}Comparison of Levinson $E^{\rm eq}_{\rm corr}$ and Born results at equal
uncorrelated kinetic energies ($K_{f}^{\rm eq}$). The initial temperature
$T_{i}$ 
of the Levinson calculation is increased to $T_{f}$ as a consequence of
the correlations. Born(Init) is the Born correlation energy at this same
temperature $T_{f}$. The remaining difference $Diff.$ between $E_{\rm
  corr}^{\rm eq}$ and $Born$ is due to the
correlational spreading of the spectral function and is discussed in the
text. All energies are in $MeV$.  The density is here normal
nuclear matter density.

\begin{tabular}[t]{|c|c|c|c|c|c|c|c|}
\hline
&  $ T_{i}$  & $E_{\rm corr}^{\rm eq}({\rm Lev})$ &$K_{f}^{i}$&$K_{f}^{\rm eq}$
&$T_{f}$&Born(Init)& $Diff.$
\\ \hline
&0&49.67&24.18&49.03&27&53.0&3.3
\\ \hline
&10&48.60&29.51&53.78&31&52.5&3.9            
\\ \hline
&20&46.74&40.91&64.24&38&51.0&4.3
\\ \hline
&40&42.14&67.93&88.98&54&46.5&4.4
\\ \hline
&60&37.22&96.40&115.00&---&---&---
\\\hline
 \end{tabular}
\end{table}

\begin{table}
%\squeezetable
\caption{\label{tab3}}Correlation energies as a function of strength of
the interaction $V_{0}$. All results are here for normal nuclear
matter density $\rho=0.183 {\rm fm}^{-3}$ and temperature $T=0$. The Born
energies are here calculated only with the initial uncorrelated
distribution.

\begin{tabular}[t]{|c|c|c|c|c|c|}
\hline
& $V_{0}$ &   $E_{\rm corr}^{\rm eq}$  (KB)  &
$E_{\rm corr}^{\rm eq}$ (Lev)&  $E_{\rm corr}^{\rm eq}$ (Born)&
$t_c$ (KB) 
\\
&MeV&MeV& MeV
&MeV&${\rm fm/c}$
\\ \hline
&906&93.15&134.41&175.9&1.8
\\ \hline
&453&35.95&49.67&43.97&2.4
\\ \hline
&227&10.45&12.41&10.99&3.2
\\ \hline
&113&2.72&2.85&2.74&3.5
\\ \hline
&57&0.68&0.70&0.69&3.6
\\ \hline
\end{tabular}
\end{table}


\begin{thebibliography}{10}
\bibitem{kad62} L.P. Kadanoff and G.Baym, Quantum Statistical Mechanics.
             (Benjamin,New York, 1962).  
\bibitem{dan84}  P. Danielewicz, Ann. \ Phys.(N.Y.) {\bf 152} 305 (1984).
\bibitem{hsk96}  H.S. K\"ohler, Phys. \ Rev. E {\bf 53} 3145 (1996).
\bibitem{LSM97} P. Lipavsk{\'y}, K. Morawetz, and V. {\v S}pi{\v c}ka,
                Annales de Physique {\bf 26},1 (2001), 
K. Morawetz, 
                Habilitation University Rostock 1998.
\bibitem{hsk95}  H.S. K\"ohler, Phys. \ Rev.C {\bf 51} 3232 (1995).
\bibitem{hsk95a}  H.S. K\"ohler, Nucl.\ Phys. {\bf A583} 339 (1995).
\bibitem{hsk95b}  H.S. K\"ohler, Proc. of the 7th International
                 Conference on Nuclear Reaction Mechanisms, Varenna, 
                 June 6-11,1994; ed.  E. Gadioli.
\bibitem{bin97}  R. Binder,H.S. K\"ohler,M. Bonitz,N.  Kwong, Phys Rev B {\bf
                 55} 5110 (1997).
\bibitem{nai98}  N.H. Kwong, M. Bonitz, R. Binder and H.S. K\"ohler,
                 Phys. Stat. Sol. {\bf 206} 197 (1998).
\bibitem{hsk97}  H.S. K\"ohler and R.Binder, Contribution to Plasma
                 Physics {\bf B37} 167 (1997).
\bibitem{bon95} M. Bonitz, D. Kremp, D.C. Scott, R. Binder,
                W. D. Kraeft and H.S. K\"ohler, J.Phys: Condens. Matter
                {\bf 8} (1996).
\bibitem{bon97} M. Bonitz, R. Binder,and H.S. K\"ohler, Contribution to
                Plasma Physics {\bf B37} 101 (1997).
\bibitem{hsk99} H.S. K\"ohler, N.H. Kwong and Hashim A. Yousif,
                Comp.Phys.Comm. {\bf 123}  123 (1999).
\bibitem{mor99}  K. Morawetz and H.S. K\"ohler, Eur. Phys. J.
                 {\bf A4} 291 (1999).
\bibitem{kra86}  W.D. Kraeft,D. Kremp,W. Ebeling and G. R\"opke,
                 "Quantum Statistics of Charged Particle Systems", 
                 Akademie-Verlag, Berlin,1986; D. Kremp,W.D. Kraeft 
                 and A.J.D. Lambert, Physica {\bf 127A} 72 (1984).
\bibitem{SLM96} V. {\v S}pi{\v c}ka, P. Lipavsk{\'y}, and K. Morawetz, 
                Phys. Lett. A {\bf 240}, 160  (1998).
\bibitem{MLSCN98} K. Morawetz {\it et~al.}, Phys. Rev. Lett. {\bf 82},
                  3767  (1999).
\bibitem{bay62} Gordon Baym and Leo P. Kadanoff,
                 Phys. \ Rev. {\bf 124} 287 (1961) ;  Gordon Baym,
                 Phys. \ Rev. {\bf 127} 1391 (1962).
\bibitem{lip86}  P. Lipavsky, V. Spicka, and B. Velicky,
                 Phys. \ Rev. B {\bf 34} 6893 (1986).
\bibitem{L65} I.~B. Levinson, Fiz. Tverd. Tela Leningrad {\bf 6},  
               2113  (1965).
\bibitem{L69} I.~B. Levinson, Zh. Eksp. Teor. Fiz. {\bf 57},  660  
              (1969), [Sov. Phys.--JETP {\bf 30}, 362 (1970)].
\bibitem{JW84} A.~P. Jauho and J.~W. Wilkins, Phys. Rev. B {\bf 29},
               1919  (1984).
\bibitem{M94}   K. Morawetz, Phys. Lett. A {\bf 199},  241  (1995).
\bibitem{MLS00} K. Morawetz, P. Lipavsk{\'y}, and V. {\v S}pi{\v c}ka, 
                Ann. of Phys.  (2000), sub. cond-mat/0005287.
\bibitem{hsk92} H.S. K\"ohler,Phys. \ Rev. C {\bf 46} 1687 (1992).
\bibitem{hsk92a} H.S. K\"ohler and Rudi Malfliet, Phys. \ Rev. C {\bf 48}
                 1034 (1992).
\bibitem{C66a} R.~A. Craig, Ann. Phys. (N.Y.) {\bf 40},  416  (1966).
\bibitem{BD68} B. Bezzerides and D.~F. DuBois, Phys. Rev. {\bf 168},
               233  (1968).
\bibitem{SZ79} H. Stolz and R. Zimmermann, Phys. Status Solidi B 
               {\bf 94},  135  (1979).
\bibitem{SR87} M. Schmidt and G. R{\"o}pke, Phys. stat. sol. (b) 
               {\bf 139},  441  (1987).
\bibitem{SL94} V. {\v S}pi{\v c}ka and P. Lipavsk{\'y}, Phys. Rev. 
                Lett {\bf 73},  3439 (1994).
\bibitem{SL95} V. {\v S}pi{\v c}ka and P. Lipavsk{\'y}, Phys. Rev. 
               B {\bf 52},  14615  (1995).
\bibitem{MLSa97} K. Morawetz, V. {\v S}pi{\v c}ka, and P. Lipavsk{\'y},
                Phys. Lett. A {\bf 246}, 311  (1998).
\bibitem{mor00}  K. Morawetz and H.S. K\"ohler,
                 Proceedings of CRIS '98
                 (2nd Catania Relativistic Ion Studies)
                 Acicastello, Italy, June 8-12, 1998, Edited by S.Costa,
                 S. Albergo, A.
                 Insolia, C. Tuve, World Scientific, Singapore, 1998.
\bibitem{BTRH92} L. Banyai, D.~B.~T. Thoai, C. Remling,  and H. Haug,
                 Phys. stat. sol. (b){\bf 173}, 149 (1992).
\bibitem{KBKS97} D. Kremp, M. Bonitz, W. Kraeft, and M. Schlanges, Ann.
                 of Phys. {\bf 258} 320 (1997).
\bibitem{P98}    S. ~L. Popyrin, Dokl. Phys. {\bf 43}, 671 (1998).
\bibitem{SPLI95} V. Spicka and P. Lipavsky, Phys.\ Rev. B {\bf 52} 14615
                 (1995).
\bibitem{Ro00}   E. P. Roth, Ph.D. thesis ,Washington University,
                 St. Louis, 2000.
\bibitem{bru55}  K.A. Brueckner, Phys.\ Rev. {\bf 100} 36 (1955). 
\bibitem{bru58}  K.A. Brueckner and J.L. Gammel, Phys.\ Rev. {\bf 109} 
                 1023 (1958).
\bibitem{So97}   H.Q. Song, M.Baldo, G.Giansiracusa, and U. Lombardo,
                 Phys. Lett. {\bf B411} (1997) 237.
\bibitem{co65}   S.A. Coon and J. Dabrowski, Phys.\ Rev. B {\bf 140} 287
                 (1965).
\bibitem{hsk73}  H.S. K\"ohler, Nucl.\ Phys. {\bf A204} 65 (1973).

\end{thebibliography}
\end{document}